\documentclass[aps,prl,nofootinbib,amsmath,amssymb,superscriptaddress,twocolumn,10pt]{revtex4}
\usepackage{graphicx}
\usepackage{dcolumn}
\usepackage{bm}
\usepackage{amssymb}
\usepackage{latexsym}
\usepackage{booktabs}
\usepackage{amsmath}
\usepackage{multirow}
\usepackage[colorlinks=true, linkcolor=red, citecolor=blue]{hyperref}

\def\be{\begin{equation}}
\def\ee{\end{equation}}
\def\ba{\begin{eqnarray}}
\def\ea{\end{eqnarray}}

\newcommand{\potential}{{A}}
\newcommand{\shift}{{B}}
\newcommand{\curvature}{{H}_L}
\newcommand{\shear}{{H}_T}
\newcommand{\kh}{k_H}
\newcommand{\ck}{c_K}
\newcommand{\tppi}{\ck p_T \Pi_T}

\begin{document}

\title{Parametrized Post-Friedmann Framework for Interacting Dark Energy}

\author{Yun-He Li}
\affiliation{Department of Physics, College of Sciences, Northeastern University, Shenyang
110004, China}
\author{Jing-Fei Zhang}
\affiliation{Department of Physics, College of Sciences, Northeastern University, Shenyang
110004, China}
\author{Xin Zhang}
\email{zhangxin@mail.neu.edu.cn}
\affiliation{Department of Physics, College of Sciences,
Northeastern University, Shenyang 110004, China}
\affiliation{Center for High Energy Physics, Peking University, Beijing 100080, China}

\begin{abstract}
Dark energy might directly interact with cold dark matter. However, in such a scenario, an early-time 
large-scale instability occurs occasionally, which may be due to the incorrect treatment for the pressure 
perturbation of dark energy as a nonadiabatic fluid. To avoid this nonphysical instability, we establish 
a new framework to correctly calculate the cosmological perturbations in the interacting dark energy 
models. Inspired by the well-known parametrized post-Friedmann approach, the condition of the 
dark energy pressure perturbation is replaced with the relationship between the momentum density of dark energy 
and that of other components on large scales. 
By reconciling the perturbation evolutions on the large and small scales, 
one can complete the perturbation equations system. The large-scale instability can be successfully avoided and the well-behaved density 
and metric perturbations are obtained within this framework. Our test results show that 
this new framework works very well and is applicable to all the interacting dark energy models.
\end{abstract}

 \maketitle

Interactions are ubiquitous in nature, and so it is very possible that dark energy directly interacts with 
cold dark matter, which also provides an intriguing mechanism to 
solve the ``coincidence problem''~\cite{coincidence,Comelli:2003cv,Zhang:2005rg,Cai:2004dk}.
The existence of such an imaginary interaction could be confirmed or falsified by the future 
highly precision measurements of the growth of large-scale structures combined with those of the expansion 
of the universe. It is also of particular importance to distinguish between the interacting dark energy (IDE) and the 
modified gravity models, since both of them can modify the growth of structures but there are some 
subtle differences between the two~\cite{Clemson:2011an}. 
To achieve this goal, the cosmological perturbations in the IDE model 
should first be investigated correctly and clearly. Numerous works on this have been done; 
see, e.g., Refs.~\cite{Clemson:2011an,Amendola:2001rc,Bertolami:2007zm,Koyama:2009gd,
Valiviita:2008iv,He:2008si}.

Nevertheless, the framework for calculating the cosmological perturbations in the IDE scenario
in the literature does not seem to be correct. This is hinted by the well-known early-time large-scale 
(superhorizon) instability appearing in the IDE scenario. 
The cosmological perturbations will blow up at the early times for the $Q\propto \rho_{c}$ model with 
$w>-1$~\cite{Valiviita:2008iv} and for the $Q\propto \rho_{de}$ model 
with $w<-1$~\cite{Clemson:2011an,He:2008si}.
Such a phenomenon is particularly prominent in the models with $w={\rm const}$. 
Here, $\rho_{de}$ and $\rho_c$ are the background energy densities of dark energy and cold dark matter, 
$w$ is the equation-of-state parameter of dark energy defined by $p_{de}=w\rho_{de}$, and $Q$ describes the interaction 
between dark energy and cold dark matter,
\begin{align}
&\rho_{de}' = -3(\rho_{de}+p_{de})+{Q_{de}\over H}\,,\label{eqn:backconsv}\\
&\rho_c' = -3\rho_c+{Q_c\over H}\,,\quad Q=Q_{de}=-Q_c\,,\label{eqn:backconsvc}
\end{align}
where $'=d/d\ln a$ and $H$ is the Hubble parameter. 
This large-scale instability seriously depresses the study of IDE. 
However, there exists an important possibility that such an instability is not a true physical effect but an 
unreal phenomenon arising from our ignorance about how to correctly treat the 
pressure perturbation of dark energy.

In general, for any adiabatic $I$ fluid its pressure perturbation takes the form $\delta p_I=c_{a,I}^2\delta\rho_I$, where $\delta\rho_I$ is the density perturbation of $I$ fluid and the adiabatic sound speed $c_{a,I}^2=p_I'/\rho_I'$. However, if dark energy is treated as an adiabatic fluid, it immediately follows that $c_{a,de}^2=w<0$ (for constant $w$ case) leading to dark energy collapsing faster than dark matter on the small scales~\cite{Gordon:2004ez}. In order to avoid such a nonphysical result, one has to treat dark energy as a nonadiabatic fluid, and lets $\delta p_{de}=c_{a,de}^2\delta\rho_{de}+ \delta p_{nad}$, where $\delta p_{nad}$ denotes the intrinsic nonadiabatic pressure perturbation of dark energy. 
As explained in Ref.~\cite{Valiviita:2008iv}, $\delta p_{nad}$ is a function of $\rho'_{de}$, and thus the interaction term $Q_{de}$ enters $\delta p_{nad}$ explicitly via Eq.~(\ref{eqn:backconsv}). Due to the interaction, the nonadiabatic mode grows fast on the large scales, no matter how small the coupling is, and soon drags other matter perturbations onto the nonadiabatic blowup, leading to rapid growth of the curvature perturbation at the early times~\cite{Valiviita:2008iv}. 
Therefore, it seems that simply treating dark energy as a fluid is problematic for IDE scenario.

In fact, even for the noninteracting dark energy such a nonadiabatic fluid treatment can also bring instability when $w$ crosses the phantom divide $w=-1$ \cite{Vikman:2004dc,Hu:2004kh,Caldwell:2005ai,Zhao:2005vj}. This instability arises due to the fact that $c_{a,de}^2=p_{de}'/\rho_{de}'$ diverges at $w=-1$. 

The appearance of these instabilities in dark energy perturbations reveals our ignorance about the nature of dark energy. 
In the current framework for calculating dark energy perturbations, dark energy is treated as a nonadiabatic fluid, and the pressure perturbation of dark energy 
is derived by assuming a rest-frame sound speed (which is not equal to the adiabatic sound speed).
But the current embarrassed situation urges us to abandon this framework and find out a new effective theory to handle dark energy perturbations.

The parametrized post-Friedmann (PPF) approach was proposed to solve the $w=-1$-crossing instability problem~\cite{Hu:2008zd,Fang:2008sn}. This approach replaces the condition on the dark energy pressure perturbation with a direct relationship between the momentum density of dark energy and that of other components on the large scales. Now, the simplified {\tt PPF} code has become a part of the {\tt CosmoMC} package~\cite{Lewis:2002ah}, used to handle the perturbations in dark energy with $w\neq{\rm const}$. 
In this letter, we establish a PPF framework for calculating the cosmological 
perturbations in the IDE scenario. In this framework, the aforementioned instability is successfully avoided. 
Also, this new PPF framework is downward compatible with the previous one for noninteracting dark energy.

In a FRW universe with scalar perturbations, the perturbed metric can be expressed in general in terms of four functions $A$, $B$, $H_L$, and $H_T$~\cite{Kodama:1985bj,Bardeen},
\begin{align}
&\delta {g_{00}} = -a^{2} (2 {\potential}Y),\qquad\delta {g_{0i}} = -a^{2} {\shift} Y_i\,,  \nonumber\\
& \qquad\delta {g_{ij}} = a^{2} (2 {\curvature} Y \gamma_{ij} + 2 {\shear Y_{ij}})\,,
\label{eqn:metric}
\end{align}
where $Y$, $Y_i$, and $Y_{ij}$ are the eigenfunctions of the Laplace operator and its covariant derivatives. 
The perturbed energy-momentum tensor can be expressed as
\begin{align}
& \delta{T^0_{\hphantom{0}0}} =  - { \delta\rho}Y,\qquad\delta{T_0^{\hphantom{i}i}} = -(\rho + p){v}Y^i\,, \nonumber\\
& \qquad \delta {T^i_{\hphantom{i}j}} = {\delta p}Y  \delta^i_{\hphantom{i}j}
	+ p{\Pi Y^i_{\hphantom{i}j}}\,,
\label{eqn:dstressenergy}
\end{align}
where $v$ and $\Pi$ are the velocity perturbation and anisotropic stress of total matters including dark energy, respectively.
Then, the Einstein equations give \cite{Hu:2008zd}
\begin{align}
& {H_L}+ {{H_T} \over 3} +   {B \over \kh}-  {H_T' \over \kh^2}\nonumber
  \\&\qquad = { 4\pi Ga^2 \over\ck k^2}   \left[ {\delta \rho} + 3  (\rho+p){{v}-
{B} \over \kh }\right] \,,\label{eqn:einstein1}\\
& {A} - {H_L'}
- { H_T' \over 3} - {K \over (aH)^2} \left( {B \over k_H} - {H_T' \over k_H^2} \right)\nonumber
  \\&\qquad =  {4\pi G \over H^2 } (\rho+p){{v}-{B} \over \kh} \,,
\label{eqn:einstein2}
\end{align}
where $k_H=k/(Ha)$ and $c_K = 1-3K/k^2$ with $k$ the wave number and $K$ the spatial curvature. Considering the interaction between dark energy and cold dark matter, the conservation laws become 
\begin{equation}
\label{eqn:energyexchange} \nabla_\nu T^{\mu\nu}_I = Q^\mu_I\,, \quad\quad
 \sum_I Q^\mu_I = 0,
\end{equation}
and the energy-momentum transfer can be split in general as 
\begin{equation}
Q_{\mu}^I  = a\big( -Q_I(1+AY) - \delta Q_IY,\,[ f_I+ Q_I (v-B)]Y_i\big),\label{eq:Qenergy}
\end{equation}
where $\delta Q_I$ and $f_I$ are the energy transfer perturbation and momentum transfer potential of $I$ fluid, respectively. Then, Eqs.~(\ref{eqn:energyexchange}) and (\ref{eq:Qenergy}) give the following two conservation equations \cite{Kodama:1985bj},
\begin{align}
& {\delta\rho_I'}
	+  3({\delta \rho_I}+ {\delta p_I})+(\rho_I+p_I)(\kh {v}_I + 3 H_L')\nonumber\\&\qquad=\frac{1}{H}(\delta Q_I-AQ_I)\,,\label{eqn:conservation1} \\
&      {[a^4(\rho_I + p_I)({{v_I}-{B}})]' \over a^4\kh}-{ \delta p_I }+ {2 \over 3}\ck p_I {\Pi_I} - (\rho_I+ p_I) {A} \nonumber\\&\qquad=\frac{a}{k}[Q_I(v-B)+f_I].\label{eqn:conservation2}
\end{align}

It is very convenient to present our work in the comoving gauge defined by $B=v_T$ and $H_T=0$. Hereafter, we use the subscript $T$ to denote the corresponding quantity of total matters except dark energy ($H_T$ is an exception). To avoid confusion, we use new symbols for metric and matter perturbation quantities in the comoving gauge. They are $\zeta\equiv H_L$, $\xi\equiv A$, $\rho\Delta\equiv\delta\rho$, $\Delta p\equiv\delta p$, $V\equiv v$, and $\Delta Q_I\equiv\delta Q_I$. Note that $\Pi$ and $f_I$ are two gauge-independent quantities. In practice, one often sets $\Pi_{de}=0$ for dark energy, $\Delta Q_{de}$ and $f_{de}$ are given by the specific interacting models, and the two metric perturbations $\zeta$ and $\xi$ satisfy two Einstein equations (\ref{eqn:einstein1}) and (\ref{eqn:einstein2}). Thus, we still have three quantities, $\rho_{de}\Delta_{de}$, $\Delta p_{de}$, and $V_{de}$, for dark energy. However, the remaining conservation equations (\ref{eqn:conservation1}) and (\ref{eqn:conservation2}) can only give two of them the equations of motion. A common practice to complete the system is to treat dark energy as a nonadiabatic fluid, and establish the relationship between $\Delta p_{de}$ and $\rho_{de}\Delta_{de}$, which, however, induces the large-scale instability in the IDE scenario, as mentioned above.

Inspired by the PPF approach to noninteracting dark energy~\cite{Hu:2008zd,Fang:2008sn}, we also replace the condition on the dark energy pressure perturbation with a direct relationship between the momentum density of dark energy and that of other components on the large scales. This relationship can be parametrized by a function $f_\zeta(a)$ as
\begin{equation}
\lim_{k_H \ll 1}
 {4\pi G \over H^2} (\rho_{de} + p_{de}) {V_{de} - V_T \over k_H}
= - {1 \over 3} \ck  f_\zeta(a) k_H V_T,\label{eq:DEcondition}
\end{equation}
since in the comoving gauge $V_{de}-V_T={\cal O}(k_H^3 \zeta)$ and $V_T={\cal O}(k_H \zeta)$ on the large scales \cite{Hu:2004xd}.
Substituting Eq.~(\ref{eq:DEcondition}) into Eq.~(\ref{eqn:einstein2}), in the comoving gauge one obtains
\begin{equation}
\lim_{k_H \ll 1} \zeta'  = \xi - {K \over k^2} k_H V_T +{1 \over 3} \ck  f_\zeta k_H V_T \,,
\label{eqn:zetaprimesh}
\end{equation}
where $\xi$ can be derived from Eq.~(\ref{eqn:conservation2}),
\begin{equation}
\xi =  -{\Delta p_T - {2\over 3}\tppi+{a\over k}[Q_c(V-V_T)+f_c] \over \rho_T + p_T}  \,.
\label{eqn:xieom}
\end{equation}

In the limit of ${k_H \gg 1}$, dark energy is smooth enough, 
and the first Einstein equation (\ref{eqn:einstein1}) reduces to the Poisson equation
\begin{equation}
\lim_{k_{H}\gg 1} \Phi = {4\pi G \over \ck k_H^2 H^2}\Delta_T \rho_T \,,
 \label{eqn:qspoisson}
\end{equation}
where $\Phi=\zeta+V_T/k_H$. In order to make these two limits compatible, one can introduce a dynamical function $\Gamma$ such that
\begin{equation}
\Phi+\Gamma = {4\pi G
\over \ck k_H^2 H^2} \Delta_T \rho_T
\label{eqn:modpoiss}
\end{equation}
on all scales. Taking the derivative of Eq.~(\ref{eqn:modpoiss}) and using Eqs.~(\ref{eqn:conservation1}), (\ref{eqn:conservation2}), and (\ref{eqn:zetaprimesh}), one can obtain the equation of motion of $\Gamma$ at $k_H \ll 1$,
\begin{equation}\label{eq:gammadot}
\lim_{k_H \ll 1} \Gamma'  = S -\Gamma \,,
\end{equation}
where the source term
\begin{align}
S&={4\pi G
\over k_H^2 H^2} \Big\{[(\rho_{de}+p_{de})-f_{\zeta}(\rho_T+p_T)]k_HV_T \nonumber\\
&\quad+{3a\over kc_K}[Q_c(V-V_T)+f_c]+\frac{1}{Hc_K}(\Delta Q_c-\xi Q_c)\Big\}.\nonumber
\end{align}
The effect of dark sector interaction is explicitly shown in this equation.

On the other hand, Eqs.~(\ref{eqn:qspoisson}), (\ref{eqn:modpoiss}), and (\ref{eq:gammadot}) imply $\Gamma\rightarrow0$ and $S\rightarrow0$ at $k_H\gg1$. To satisfy all these limits at $k_H\ll1$ and $k_H\gg1$, we can take the equation of motion for $\Gamma$ on all scales to be
\begin{equation}
(1 + c_\Gamma^2 k_H^2) [\Gamma' + \Gamma + c_\Gamma^2 k_H^2 \Gamma] = S\,.
\label{eqn:gammaeom}
\end{equation}
Here, $c_\Gamma$ gives a transition scale in terms of the Hubble scale under which dark energy is smooth enough. Once the evolution of $\Gamma$ is derived, we can directly obtain the energy density perturbation and momentum density of dark energy,
\begin{align}
&\rho_{de}\Delta_{de} =- 3(\rho_{de}+p_{de}) {V_{de}-V_{T}\over k_{H} }-{k^{2}\ck \over 4\pi G a^{2}} \Gamma \,,\\ \label{eqn:ppffluid}
&{ V_{de}-V_{T}\over k_{H}} ={-H^{2} \over 4\pi G (\rho_{de} + p_{de}) F} \nonumber \\
&\quad\quad\quad\times\left[ S - \Gamma' - \Gamma + f_{\zeta}{4\pi G (\rho_{T}+p_{T}) \over H^{2}}{V_{T}\over k_{H}}
\right]  \,,
\end{align}
with $$F = 1 + 3 {4 \pi G a^2 \over k^2 \ck} (\rho_T + p_T).$$ 
Actually, the dark energy pressure perturbation $\Delta p_{de}$ can also be derived within this framework from Eq.~(\ref{eqn:conservation2}),
\begin{align}
&\Delta p_{de}={[a^{4}(\rho_{de}+p_{de})(V_{de}-V_T)]' \over a^{4}k_{H}} - (\rho_{de}+p_{de}) \xi \nonumber\\
&\qquad~-{Q_{de}\over H}{(V-V_T)\over k_H}-{f_{de}\over k_H} \,.
\label{eq.dpde}
\end{align}

So far, the perturbation system has been completed by a function $f_\zeta$, a parameter $c_\Gamma$, and a dynamical quantity $\Gamma$. We can take the initial condition $\Gamma=0$ at $a\rightarrow0$ for solving Eq.~(\ref{eqn:gammaeom}), since $S\rightarrow0$ at $a\rightarrow0$ from the expression of $S$. The values of $f_\zeta$ can be inferred by solving the full equations at $k_H\rightarrow0$ in a specific IDE model. However, it suffices for most purposes to simply let $f_\zeta=0$ \cite{Fang:2008sn}. For the value of $c_\Gamma$, we follow Ref.~\cite{Fang:2008sn} and choose it to be $0.4c_s$~\cite{cs1,cs2}. 
With a careful test, we conclude that the dark energy perturbation evolution is insensitive to this value.

Next, we show that this new framework can give stable cosmological perturbations in the IDE scenario. 
As a concrete example, we consider the following typical model,
\begin{equation}\label{eq:covQ}
Q^{\mu}_c  = -Q^{\mu}_{de}=-3\beta H\rho_c u^{\mu}_c,
\end{equation}
where $\beta$ is a dimensionless coupling. The four-velocities for $I$ fluid in a general gauge are $$u^\mu_I = a^{-1}\big(1-AY,\,v_IY^i \big)\,,\quad u_\mu^I = a\big(-1-AY,\,(v_I -B)Y_i\big).$$
From Eqs.~(\ref{eq:Qenergy}) and (\ref{eq:covQ}), we have
 \begin{align}
&\delta Q_{de}= -\delta Q_c=3\beta H\rho_c \delta_c,\quad f_{de}=-f_c=3\beta H\rho_c (v_c-v),\nonumber \\
 &\qquad\qquad\qquad Q_{de}=-Q_c=3\beta H\rho_c,
 \end{align}
where $\delta_I=\delta\rho_I/\rho_I$ denotes the dimensionless density perturbation of $I$ fluid.
Substituting the above equations into the source term $S$ and Eq.~(\ref{eqn:xieom}), we can obtain the perturbations of dark energy in the comoving gauge. It is also convenient to get the results in the synchronous gauge by a gauge transformation, since most public numerical codes are written in the synchronous gauge. For the details of the gauge transformation, see the appendix of Ref.~\cite{Hu:2008zd} (but note that the background interaction term $Q_I$ enters the transformations of $\delta_I$ and $\delta p_I$ via Eqs.~(\ref{eqn:backconsv}) and (\ref{eqn:backconsvc})).

\begin{table}[htbp]\caption{Fit results for the IDE model with $Q^\mu=3\beta H\rho_c u_c^\mu$.}\label{tab.params}
\begin{tabular}{lcc}
\hline\hline
Parameter  & Best fit & 68\% limits \\
\hline
$\Omega_bh^2$&0.02227&$0.02218\pm0.00028$\\
$\Omega_ch^2$&0.12199&$0.1224\pm0.0022$\\
$H_0$&71.16&$71.5\pm1.5$\\
$\tau$&0.0955&$0.090^{+0.012}_{-0.014}$\\
$w$&$-1.2050$&$-1.228^{+0.093}_{-0.084}$\\
$\beta$&$-0.00137$&$-0.0013\pm0.0008$\\
$n_s$&0.9630&$0.9609^{+0.0066}_{-0.0065}$\\
${\rm{ln}}(10^{10}A_{\rm{s}})$&3.096&$3.086^{+0.024}_{-0.027}$\\
\hline
$\Omega_\Lambda$&0.7139&$0.7152^{+0.0128}_{-0.0116}$\\
$\Omega_m$&0.2861&$0.2848^{+0.0116}_{-0.0128}$\\
${\rm{Age}}/{\rm{Gyr}}$&13.826&$13.831\pm0.064$\\
\hline
\end{tabular}
\end{table}

\begin{figure*}[!htbp]
  \includegraphics[width=7cm]{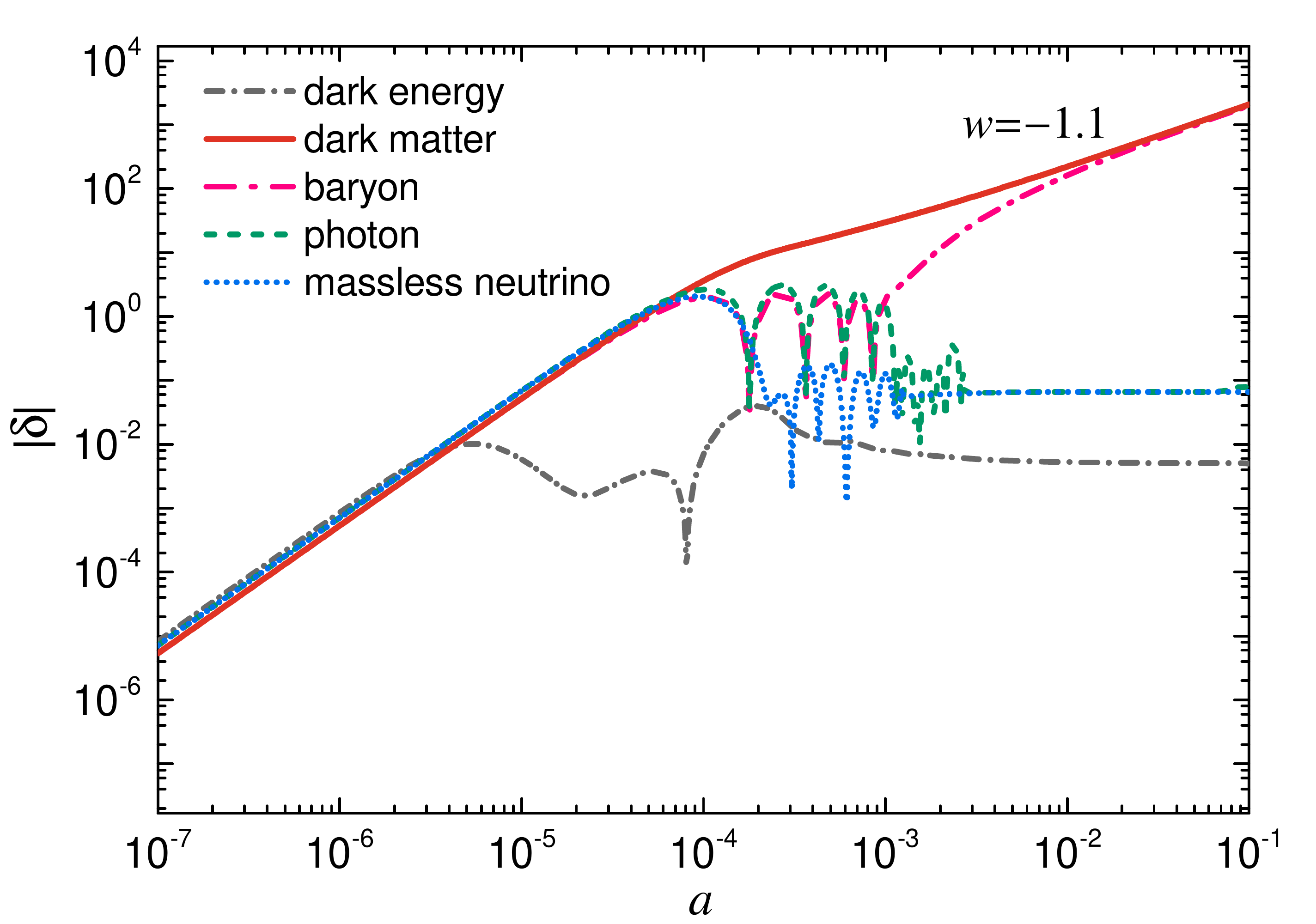}
  \includegraphics[width=7cm]{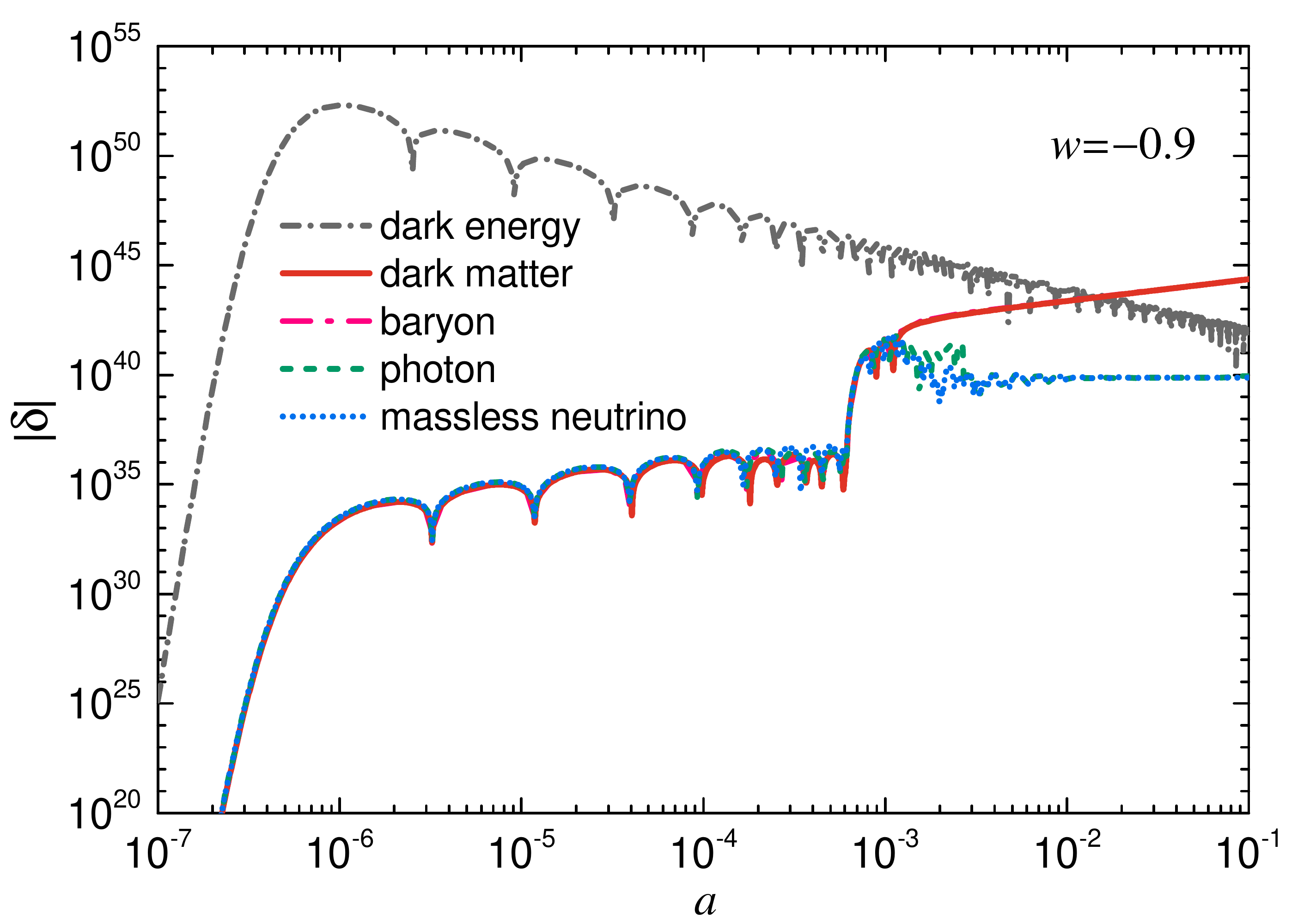}
  \includegraphics[width=7cm]{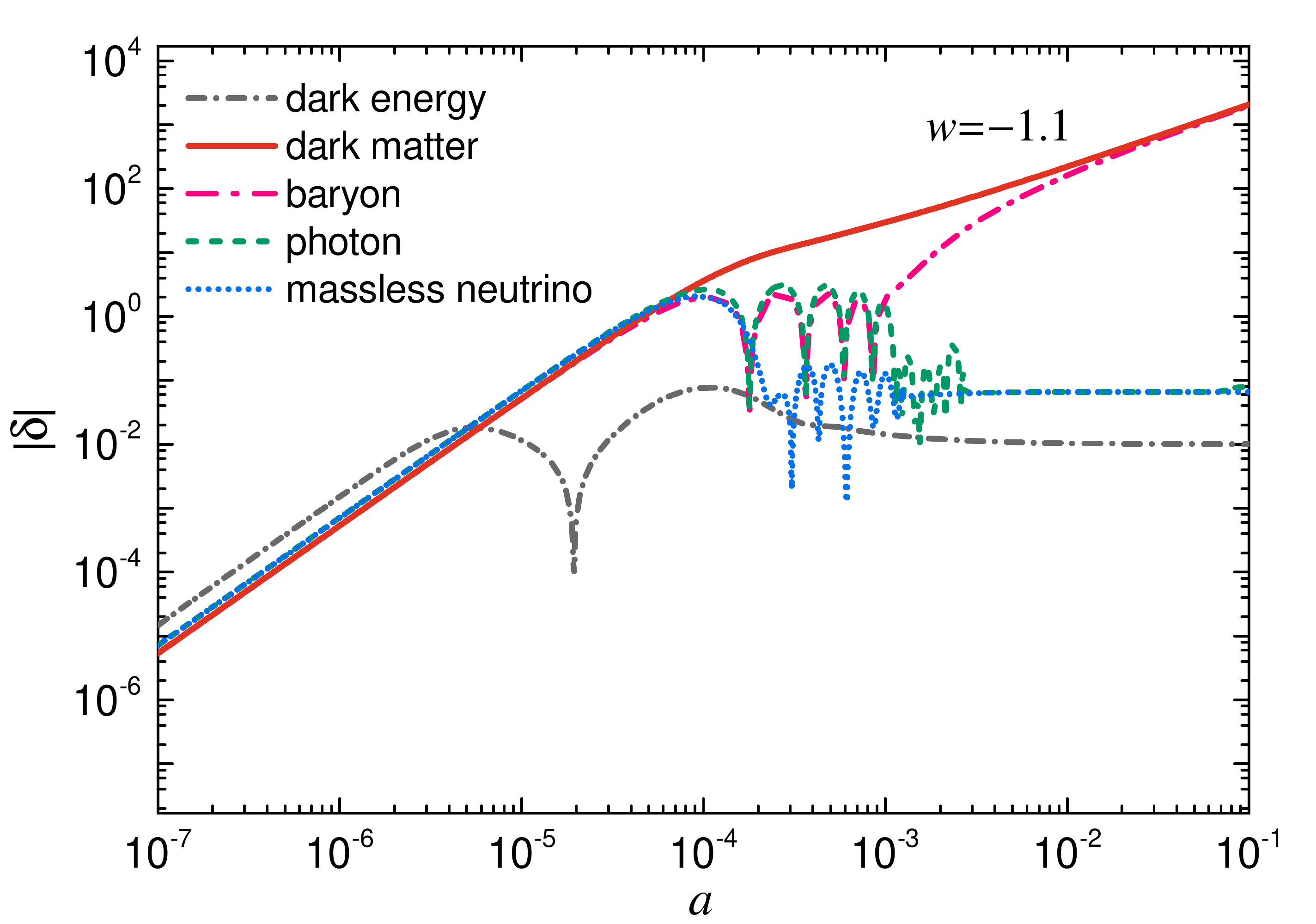}
  \includegraphics[width=7cm]{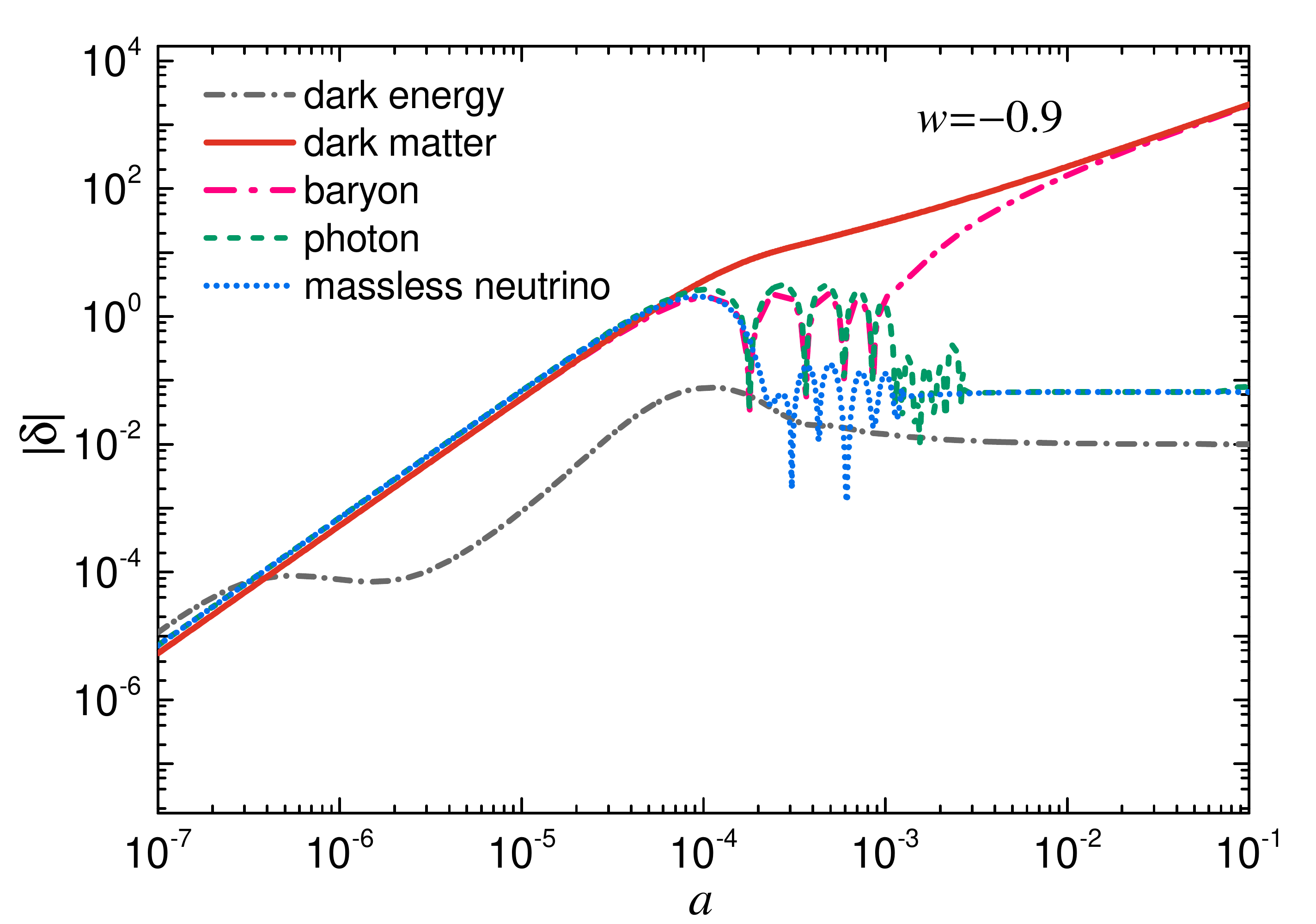}
  \caption{The density perturbation evolutions at $k=0.1~\rm{Mpc}^{-1}$ in the IDE model with $Q^\mu=3\beta H\rho_c u_c^\mu$ in the synchronous gauge. The upper panels are obtained by using the previous method, while the lower panels are obtained within the PPF framework proposed in this work. 
}\label{perturbationevolve}
\end{figure*}

\begin{figure}[!htbp]
  \includegraphics[width=8.5cm]{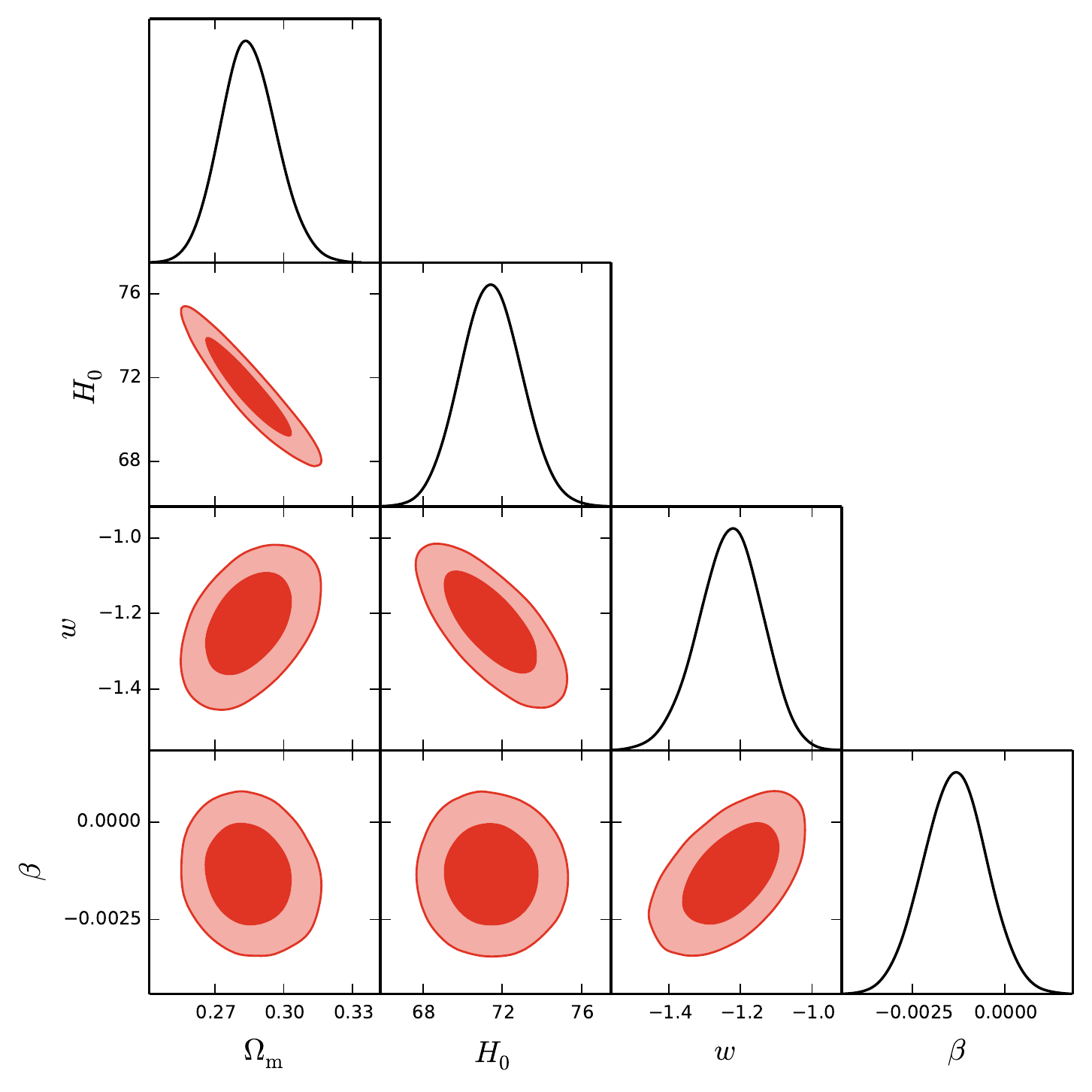}
  \caption{The one- and two-dimensional posterior distributions for the parameters 
  in the IDE model with $Q^\mu=3\beta H\rho_c u_c^\mu$.}\label{fig.fitresult}
\end{figure}

In Fig.~\ref{perturbationevolve}, we plot the density perturbation evolutions at $k=0.1~\rm{Mpc}^{-1}$ for the considered IDE model in the synchronous gauge. The upper panels are obtained by using the previous method, while the lower panels are obtained within the PPF framework proposed in this letter. 
To show the cases with $w<-1$ and $w>-1$, we take $w=-1.1$ (left panels) and $w=-0.9$ (right panels) 
as typical examples. 
We take $\beta=-10^{-17}$ and fix all the other parameters at their best-fit values from Planck. 
Taking such a small value for $\beta$ is to avoid the possible breakdown of the numerical computation 
when the instability occurs in the IDE model using the old method.
From Fig.~\ref{perturbationevolve}, one can clearly see that our new calculation framework (lower panels) can give stable cosmological perturbations for both $w=-1.1$ and $w=-0.9$ cases, while the previous method (upper panels) leads to the instability for the $w=-0.9$ case even though the coupling is so weak ($\beta=-10^{-17}$). Thus, our new calculation scheme works very well. Note that Fig.~\ref{perturbationevolve} is only an example for the IDE model with $Q^\mu=3\beta H\rho_c u_c^\mu$. In fact, after a careful test, we conclude that our new framework is 
applicable to all the IDE models.

Also as an example, we constrain the parameter space for the above IDE model by using the current observations. 
The observational data are the same as those used in Ref.~\cite{Li:2013bya}. 
The fit results are shown in Table~\ref{tab.params} and Fig.~\ref{fig.fitresult}. 
This example explicitly shows that the whole parameter space of the IDE model can be explored within this new 
calculation framework. In this fit, we get $\beta=-0.0013\pm0.0008$ and $w=-1.228^{+0.093}_{-0.084}$ 
(1$\sigma$ CL).

Dark energy might interact with cold dark matter in a direct, nongravitational way. 
The consideration of such an interaction is rather natural since the interactions are ubiquitous in nature. 
On the contrary, no interaction between dark energy and dark matter 
is an additional assumption~\cite{ref:Peebles2010}. 
In order to find out this interaction and determine the properties of dark energy and dark matter with 
the future highly accurate measurements of the growth of large-scale structure, one should investigate 
the cosmological perturbations in detail in the IDE scenario. 
However, some early-time large-scale instability occurs on occasion in the IDE scenario, due to the 
incorrect treatment for the pressure perturbation of dark energy as a nonadiabatic fluid. 
In this letter, we establish a PPF framework to 
correctly calculate the cosmological perturbations in the IDE scenario, in which the dark energy pressure 
perturbation condition is replaced by the relationship between the momentum density of dark energy and 
that of the other components on large scales (parametrized by a function $f_\zeta$). 
Using a dynamical quantity $\Gamma$ and a transition-scale parameter $c_\Gamma$ to reconcile the perturbation evolutions on the small and large scales, 
the density and velocity perturbations of dark energy can be 
derived directly.
Our new framework can give stable cosmological perturbations in the whole expansion 
history of the universe, and is applicable to all the IDE models. 
This calculation scheme would play a crucial role in distinguishing among the (noninteracting) dark energy, 
IDE, and modified gravity models with future highly precision data, and inject new vitality to the study of IDE models.

\begin{acknowledgments}
We acknowledge the use of {\tt CosmoMC}. This work was supported by the National Natural Science Foundation of China (Grant No.~11175042) and the Fundamental Funds for the Central Universities (Grant No.~N120505003).
\end{acknowledgments}


\begin{thebibliography}{99}
\bibitem{cs1}
Here, $c_s$ specifically denotes the sound speed of a scalar-field dark energy. 
Note that the PPF approach itself does not need to define a dark energy sound speed $c_s$. 
However, in our PPF framework, the dark energy pressure perturbation actually can also be 
derived [see Eq.~(\ref{eq.dpde})], and so one can immediately obtain the value of the effective sound speed by its definition. 
In fact, the PPF approach can exactly match with the usual fluid approach for the stable large-scale dark energy perturbation by calibrating the function $f_\zeta$ in Eq.~(\ref{eq:DEcondition}), 
and thus assumes $c_s^2=1$ on the large scales. 
On the small scales, the PPF approach actually assumes $c_s^2\gg1$, since in the 
limit of $k_H\gg 1$ dark energy is assumed to be smooth within the PPF framework.
\bibitem{cs2}
In addition, the effective rest-frame sound speed derived in the PPF framework must depend on the background coupling with cold dark matter. 
This is natural and physical, since once dark energy interacts with dark matter, the dark energy fluid cannot be viewed to only depend on dark energy. 
In fact, the dark energy fluid is now a mixture of dark energy and dark matter due to the coupling between them. 
Thus, it is obvious that the sound speed of dark energy is determined by the microphysics of both dark energy and dark matter. 
For a similar example, recall the baryon sound speed in the case of baryons coupling to photons.
\end{thebibliography}
\end{document}